\documentclass[aps, prb, amsmath, amssymb, amsfonts, twocolumn, showpacs]{revtex4}
\usepackage{bm,graphicx,color}

\newcommand{\be}{\begin{equation}}
\newcommand{\bea}{\begin{eqnarray}}
\newcommand{\ee}{\end{equation}}
\newcommand{\eea}{\end{eqnarray}}

\newcommand{\bc}{\begin{center}}
\newcommand{\ec}{\end{center}}

\begin{document}

\title{Low-energy properties of the ferromagnetic metallic phase
in manganites: Slave fermion approach to the quantum double exchange
model}
\author{Yu-Li Lee}
\affiliation{Physics Department, National Changhua University of Education, Changhua, Taiwan, R.O.C.}
\author{Yu-Wen Lee}
\affiliation{Physics Department, Tunghai University, Taichung, Taiwan, R.O.C.}
\date{\today}
\begin{abstract}
We study the low energy properties of the one-orbital quantum
double-exchange model by using the slave fermion formulation. We
construct a mean-field theory which gives a simple explanation for
the magnetic and thermodynamic properties of the ferromagnetic
metallic phase in manganites at low energy. The resulting electron
spectral function and tunneling density of states show an incoherent
asymmetric peak with weak temperature dependence, in addition to a
quasiparticle peak. We also show that the gauge fluctuations in the
ferromagnetic metallic phase are completely screened due to the
Anderson-Higgs mechanism. Therefore, the mean-field state is robust
against gauge fluctuations and exhibits spin-charge separation at
low energy.
\end{abstract}
\pacs{75.47.Lx, 75.10.Lp, 79.60.-i}

\maketitle

\section{Introduction}

The phenomenon of colossal magnetoresistance (CMR) is known to occur
in various doped perovskite manganese oxides with the chemical
formula Re$_{1-x}$A$_x$MnO$_3$, where Re is the rare earth such as
La or Nd and A is a divalent alkali such as Sr or Ca.\cite{Sch} More
recent studies revealed a complex phase diagram and very rich
physics.\cite{Dagotto} Therefore, it is of great theoretical
interest to find a proper minimal model for these CMR systems, which
can account for the important common features, such as the transport
and magnetic properties, shared by all these materials, while
leaving out many non-universal peculiarities due to crystal
environment and atomic structure of individual compounds.

One of the universally recognized common feature of these CMR
compounds is the sizable ferromagnetic (FM) Hund's rule coupling
$J_H$ between the core spins and those of the $e_g$ electrons. This
gives the double-exchange (DE)
interaction\cite{Zener,Anderson,deG,spin-wave-zeroU,Furukawa} which
is believed to be a fundamental mechanism in explaining many of the
interesting features of the CMR compounds. From this point of view,
a usual starting point to analyze the properties of the CMR material
is the standard DE Hamiltonian, supplemented with other terms, such
as the Hubbard repulsion $U$, the antiferromagnetic (AF)
superexchange interactions between core spins $J_{AF}$, etc. For the
case of CMR manganites, band theory calculations\cite{band} suggest
the typical values of the hopping amplitude $t\sim 0.3-0.5$ eV,
$J_H\sim 2.5$ eV, and $U\sim 6-8$ eV.

Motivated by experimental findings, there has been significant
progress in the study of the DE Hamiltonian, notably applying the
Schwinger boson\cite{Sarkar, Khaliullin2} or $1/S_c$
expansion.\cite{Golosov2, Shannon} ($S_c=3/2$ is the core spin.)
Most of these approaches lead to a simple Fermi liquid (a doped
band insulator) picture for the ground state of the FM metallic
phase, which cannot fully explain the results of experiments,
especially the optical conductivity at low
energy\cite{Okimoto,Kim} and the angle-resolved photoemission
(ARPES) measurements.\cite{Dessau, Chuang,Sarma} Recently, Golosov
tried to address parts of the discrepancy by including the on-site
Coulomb repulsion through the Hartree-Fock
approximation.\cite{Golosov} Although some interesting results
were found, the Hartree-Fock approximation is still based on the
Fermi-liquid picture. To capture the low-energy physics for large
values of $U$ and $J_H$ properly, a slave fermion approach has
been proposed.\cite{Weisse, Khaliullin2} A recent work by
Hu\cite{Hu} along this direction has shown that the massless
fluctuations of the longitudinal part of the gauge fields arising
from the slave-fermion approach indeed dramatically change the
behavior of the spectral function of $e_g$ electrons at low
energy.

The purpose of the present article is to give a simple mean-field
description on the low temperature properties of the DE system in
the large $U,J_H$ limit. (For the implementation of the large
$U,J_H$ limit, see Sec. \ref{model}.) Orbital fluctuations,
Jahn-Teller effect, and nanoscale phase separation, though very
interesting, make analysis difficult. Therefore, we shall focus our
attention on the region where the DE mechanism is the dominant
factor, namely, the region at the hole concentration $0.2<x<0.5$ and
away from the critical temperature $T_c$. We propose a mean-field
theory based on the slave fermion scheme, which exhibits spin-charge
separation at low energy, and use it to calculate various low-energy
properties in the FM metallic phase. Among them, the most important
result is that the quasiparticle peak in the electron spectral
function is reduced at low temperatures and the spectral weight is
transferred to an asymmetric broad peak away from the Fermi surface,
which is a natural consequence of the spin-charge separation in our
theory. The other physical properties in the FM metallic phase we
have studied are as follows: (i) The magnon dispersion at the
mean-field level is similar to that of a simple cubic Heisenberg
ferromagnet, which has been verified by inelastic neutron scattering
measurements. (ii) The magnitudes of the spin stiffness at $T=0$ and
the coefficients in the low-temperature specific heat, obtained from
the mean-field theory, are consistent with experimental data. (iii)
The structure of the optical conductivity at low energy is of the
form similar to that observed by experiments, and the Drude weight
is reduced.

We now briefly outline the structure of our paper. In Sec.
\ref{model}, we will introduce the quantum double exchange (QDE)
model and develop a slave fermion mean-field theory. The results
of the mean-field theory is presented in Sec. \ref{result}. We
will show the stability of our mean-field state against the gauge
fluctuations in Sec. \ref{gauge}. The last section is devoted to a
conclusive discussion.

\section{The model and the slave fermion mean-field theory}
\label{model}

We shall start with the quantum double-exchange (QDE) model
described by the following Hamiltonian:\cite{Weisse}
\begin{equation}
 H=-t\sum_{\bm{i},\bm{u},\sigma}\left(\bar{c}^{\dagger}_{\bm{i}+\bm{u}\sigma}
   \bar{c}_{\bm{i}\sigma}+{\mathrm H.c.}\right)-J_H\sum_{\bm{i}}\bm{S}_{\bm{i}}^c
   \cdot \bm{s}_{\bm{i}} \ , \label{qdeh1}
\end{equation}
where the first and second terms describe electronic hopping and
DE couplings, respectively. Here, $\bm{S}^c$ denotes the core
spin, $\bm{s}$ is the spin operator of $e_g$ electrons, $\bm{u}$
is a unit vector connecting the nearest-neighbor sites around site
$\bm{i}$ (for a simple cubic lattice, $\bm{u}=\hat{\bm{x}}$,
$\hat{\bm{y}}$, and $\hat{\bm{z}}$), and $\bar{c}_{\bm{i}\sigma}$
is the annihilation operator of $e_g$ electrons at site $\bm{i}$
with spin $\sigma$. In Eq. (\ref{qdeh1}), we have neglected the
superexchange interactions between core spins. This is because
$J_{AF}\sim 5-10$ meV in manganites, which is much smaller than
$t$. Moreover, in the undoped compounds, a Jahn-Teller distortion
lifts the orbital degeneracy of the $e_g$ electrons with the
energy scale $E_{JT}\sim 1-1.6$ eV. Thus, to study the low energy
physics in the hole-doped region ($x<0.5$), one may in a first
approximation ignore the orbital degeneracy and apply the
one-orbital model. To take into account the fact that the on-site
Coulomb interactions between $e_g$ electrons is the largest energy
scale in manganites, we impose the no-double-occupancy (NDO)
condition on the $e_g$ electron operators:
\begin{equation}
 n_{\bm{i}}\equiv \sum_{\sigma}\bar{c}^{\dagger}_{\bm{i}\sigma}
   \bar{c}_{\bm{i}\sigma}=0,1 \ . \label{ndo1}
\end{equation}
Therefore, $\bar{c}_{\bm{i}\sigma}$ and
$\bar{c}^{\dagger}_{\bm{i}\sigma}$ cannot obey the canonical
anticommutation relations, and Eq. (\ref{qdeh1}) is valid for
energies much lower than $U$.

Because of the NDO condition [Eq. (\ref{ndo1})], the system becomes
a strongly correlated one. One popular way to solve this kind of
problems is to introduce a pair of slave fields to rewrite the
$\bar{c}$ operator. Here we adopt the slave-fermion representation:
\begin{equation}
 \bar{c}_{\bm{i}\sigma}=f^{\dagger}_{\bm{i}}b_{\bm{i}\sigma} \ ,
     \label{sfr1}
\end{equation}
where $f_{\bm{i}}$ and $f_{\bm{i}}^{\dagger}$ satisfy the canonical
anticommutation relations, and $b_{\bm{i}\sigma}$ and
$b_{\bm{i}\sigma}^{\dagger}$ satisfy the canonical commutation
relations.\cite{sfbook} In this way, the charge and spin degrees of
freedom of $e_g$ electrons are represented by $f$ (holon) and $b$
(spinon) fields, respectively. In terms of $f_{\bm{i}}$ and
$b_{\bm{i}\sigma}$, the NDO condition becomes an identity
\begin{equation}
 \sum_{\sigma}b^{\dagger}_{\bm{i}\sigma}b_{\bm{i}\sigma}
     +f^{\dagger}_{\bm{i}}f_{\bm{i}}=1 \ . \label{ndo2}
\end{equation}

In manganites, one may further simplify Eq. (\ref{qdeh1}). Since
$J_H\sim 10t$, at the energy scale much lower than $J_H$, it
suffices to consider the Hilbert space in which the $e_g$ electron
spin is parallel to the core spin. One may introduce another
Schwinger bosons $d_{\bm{i}\sigma}$ and $d^{\dagger}_{\bm{i}\sigma}$
to describe the total spin. Within this subspace, it can be shown
that the $b$ field is associated with the $d$ field through the
relation\cite{Khaliullin2,Weisse}
\begin{equation}
 b_{i\sigma}=\frac{1}{\sqrt{2S}}d_{i\sigma} \ , \label{sb3}
\end{equation}
with $S=2$. Collecting the above results, in the limit
$J_H/t\rightarrow +\infty$, Eq. (\ref{qdeh1}) is reduced to the one
\begin{equation}
 H_{DE}=-t\sum_{\bm{i},\bm{u},\sigma}\left(\bar{c}^{\dagger}_{\bm{i}+\bm{u}\sigma}
  \bar{c}_{\bm{i}\sigma}+{\mathrm H.c.}\right) \ , \label{qdeh2}
\end{equation}
and the NDO condition becomes
\begin{equation}
 \sum_{\sigma}d^{\dagger}_{\bm{i}\sigma}d_{\bm{i}\sigma}
     +f^{\dagger}_{\bm{i}}f_{\bm{i}}=2S \ . \label{ndo3}
\end{equation}
We shall take Eq. (\ref{qdeh2}) as a minimal model to describe the
FM metallic phase in manganites.\cite{foot1}

To proceed, we turn into the path-integral formalism. The
partition function of the QDE model in the large $J_H$ limit can
be written as
\begin{equation}
 Z=\int \! \! D[f^{\dagger}]D[f]D[d_{\sigma}^{\dagger}]
      D[d_{\sigma}]D[\lambda]\exp{ \! \left\{- \! \int^{\beta}_0 \! \!
      d\tau \sum_{\bm{i}}L\right\}} , \label{qdez1}
\end{equation}
where
\begin{eqnarray*}
 L &=& \sum_{\sigma}d^{\dagger}_{\bm{i}\sigma}(\partial_{\tau}+\lambda_{\bm{i}})
   d_{\bm{i}\sigma}+f^{\dagger}_{\bm{i}}(\partial_{\tau}+i\lambda_{\bm{i}}-\mu_0)
   f_{\bm{i}} \\
   & & +\frac{t}{2S}\sum_{\bm{u},\sigma}\left(f^{\dagger}_{\bm{i}+\bm{u}}
   f_{\bm{i}}d^{\dagger}_{\bm{i}\sigma}d_{\bm{i}+\bm{u}\sigma}+{\mathrm H.c.}
   \right) \! -2iS\lambda_{\bm{i}} \ ,
\end{eqnarray*}
$\mu_0$ is the chemical potential of holes, and $\lambda_{\bm{i}}$
is the Lagrangian multiplier to impose the NDO condition. To
facilitate the mean-field analysis, one may perform the
Hubbard-Stratonovich transformation to decouple the hopping term,
and the Lagrangian becomes
\begin{eqnarray}
 L &=& \sum_{\sigma}d^{\dagger}_{\bm{i}\sigma}(\partial_{\tau}+i
   \lambda_{\bm{i}})d_{\bm{i}\sigma}+f^{\dagger}_{\bm{i}}
   (\partial_{\tau}+i\lambda_{\bm{i}}-\mu_0)f_{\bm{i}} \nonumber
   \\
   & & -\frac{t}{2S}\sum_{\bm{u}} \! \left(\chi^{\dagger}_{\bm{i}+\bm{u},\bm{i}}
   f^{\dagger}_{\bm{i}+\bm{u}}f_{\bm{i}}+{\mathrm H.c.}\right)
   \nonumber \\
   & & -\frac{t}{2S}\sum_{\bm{u},\sigma} \! \left(\eta^{\dagger}_{\bm{i}+\bm{u},\bm{i}}
   d^{\dagger}_{\bm{i}+\bm{u}\sigma}d_{\bm{i}\sigma}+{\mathrm H.c.}
   \right) \nonumber \\
   & & -\frac{t}{S}\sum_{\bm{u}} \! \left(\eta^{\dagger}_{\bm{i}+\bm{u},\bm{i}}
   \chi_{\bm{i}+\bm{u},\bm{i}}+{\mathrm H.c.}\right) \! -2iS
   \lambda_{\bm{i}}, \label{qdell1}
\end{eqnarray}
We notice that there is a U(1) gauge structure in the slave-fermion
scheme, which is reflected in the invariance of $L$ (up to a total
derivative in $\tau$) under the U($1$) gauge transformation:
\begin{eqnarray}
 & & f_{\bm{i}}\rightarrow f_{\bm{i}}e^{-iw_{\bm{i}}} \ , ~~
     d_{\bm{i}\sigma}\rightarrow d_{\bm{i}\sigma}e^{-iw_{\bm{i}}}
     \ , \nonumber \\
 & & \chi_{\bm{i}+\bm{u}, \bm{i}}\rightarrow \chi_{\bm{i}+\bm{u}, \bm{i}}
     e^{i(w_{\bm{i}+\bm{u}}-w_{\bm{i}})} \ , \nonumber \\
 & & \eta_{\bm{i}+\bm{u}, \bm{i}}\rightarrow \eta_{\bm{i}+\bm{u}, \bm{i}}
     e^{i(w_{\bm{i}+\bm{u}}-w_{\bm{i}})} \ , \nonumber \\
 & & \lambda_{\bm{i}}\rightarrow \lambda_{\bm{i}}+\partial_{\tau}
     w_{\bm{i}} \ , \label{u1gauge11}
\end{eqnarray}

Motivated by the DE mechanism, which allows the coherent hopping of
charge carriers in a FM background, we consider the following
mean-field ansatz:
\begin{equation}
 \chi_{\bm{i}+\bm{u},\bm{i}}=\chi \ , ~~
 \eta_{\bm{i}+\bm{u},\bm{i}}=\eta \ , ~~
 i\lambda_{\bm{i}}=\Delta \ , \label{urvb31}
\end{equation}
where $\chi$, $\eta$, and $\Delta$ are real. Inserting Eq.
(\ref{urvb31}) into Eq. (\ref{qdell1}), the mean-field Lagrangian
can be written as
\begin{eqnarray*}
 L_{mf} &=& \sum_{\sigma}d^{\dagger}_{\bm{i}\sigma}(\partial_{\tau}
   +\Delta)d_{\bm{i}\sigma}+f^{\dagger}_{\bm{i}}(\partial_{\tau}
   +\Delta-\mu_0)f_{\bm{i}} \\
   & & -\frac{\chi t}{2S}\sum_{\bm{u}} \! \left(f^{\dagger}_{\bm{i}+\bm{u}}
   f_{\bm{i}}+{\mathrm H.c.}\right) \\
   & & -\frac{\eta t}{2S}\sum_{\bm{u},\sigma} \! \left(
   d^{\dagger}_{\bm{i}+\bm{u}\sigma}d_{\bm{i}\sigma}+{\mathrm H.c.}
   \right) \\
   & & -\frac{zt}{S}\eta \chi -2S\Delta \ ,
\end{eqnarray*}
where $z=6$ is the coordination number. Now $L_{mf}$ becomes
quadratic in $f$ and $d_{\sigma}$, and one may integrate them out to
obtain the mean-field grand potential. The mean-field equations are
obtained through minimizing the mean-field grand potential with
respect to the mean-field parameters, yielding
\begin{eqnarray}
 \frac{1}{N}\sum_{\bm{k}}n_B(\epsilon_{b\bm{k}}+\Delta) &=& S-\frac{x}{2}
      \ , \label{qdemfe1} \\
 \frac{1}{N}\sum_{\bm{k}}\epsilon_{f\bm{k}}n_F(\epsilon_{f\bm{k}}-\mu_f)
      &=& \frac{zt}{S}\chi \eta \ , \label{qdemfe2} \\
 \frac{1}{N}\sum_{\bm{k}}\epsilon_{b\bm{k}}n_B(\epsilon_{b\bm{k}}+\Delta)
      &=& \frac{zt}{2S}\chi \eta \ , \label{qdemfe3}
\end{eqnarray}
where $n_F(x)=(e^{\beta x}+1)^{-1}$ is the Fermi-Dirac distribution,
$n_B(x)=(e^{\beta x}-1)^{-1}$ is the Bose-Einstein distribution, $N$
is the number of lattice points, $\epsilon_{f\bm{k}}=-\frac{\chi
t}{S}\sum_{\bm{u}}\cos{(\bm{k}\cdot \bm{u})}$, and
$\epsilon_{b\bm{k}}=-\frac{\eta t}{S}\sum_{\bm{u}}\cos{(\bm{k}\cdot
\bm{u})}$. The chemical potential of holons, $\mu_f=\mu_0-\Delta$,
is associated with the hole concentration $x$ through by the
equation
\begin{equation}
 \frac{1}{N}\sum_{\bm{k}}n_F(\epsilon_{f\bm{k}}-\mu_f)=x \ .
      \label{qdehn1}
\end{equation}
Equations (\ref{qdemfe1}) --- (\ref{qdehn1}) are the mean-field
equations we want to solve.

Before solving these mean-field equations numerically, we may look
into the physics revealed by them. Equation (\ref{qdemfe1}) is
nothing but the number equation of free bosons with the average
number of particles per site $S-x/2$ and the chemical potential
$-\Delta$. In three dimensions, there exists a critical temperature
$T_c>0$ so that $\Delta =\Delta_0$ for $T\leq T_c$ and $\Delta
>\Delta_0$ for $T>T_c$, where $\Delta_0=\eta zt/(2S)$ is the
bottom of the spinon band $\epsilon_{b\bm{k}}$. When $\Delta
=\Delta_0$, the Bose-Einstein condensation (BEC) of the $d$ bosons
occurs, which corresponds to the FM phase. On the other hand, it
is the paramagnetic (PM) phase for $\Delta >\Delta_0$. In this
scenario, the PM to FM phase transition in manganites is
associated with the BEC of the $d$ bosons, and $T_c$ corresponds
to the Curie temperature.

With the above understanding in mind, we can see that deep inside
the FM phase, without loss of generality, one may choose the
direction of magnetization to be the $z$ axis and set $\langle
d_{\bm{i}\uparrow}\rangle =\sqrt{2S-x}$ and $\langle
d_{\bm{i}\downarrow}\rangle =0$ in terms of a proper spin SU($2$)
rotation. (This parametrization for $d$ bosons is consistent with
the experiment which shows the complete spin polarization of
conduction electrons in manganites.\cite{Park2}) Further, for $T\ll
T_c$, one may neglect the amplitude fluctuations of
$d_{\bm{i}\uparrow}$. As for the phase fluctuations of
$d_{\bm{i}\uparrow}$, it can be absorbed into $d_{\bm{i}\downarrow}$
and $f_{\bm{i}}$ by choosing a particular gauge or performing a
proper U($1$) gauge transformation with the help of the U($1$) gauge
invariance of $L$. Thus, to obtain the low-energy effective
Hamiltonian in the FM phase, it suffices to set
\begin{equation}
 d_{\bm{i}\uparrow}=\sqrt{2S-x} \ , ~~ d_{\bm{i}\downarrow}=b_{\bm{i}}
   \ , \label{qdemf1}
\end{equation}
in $L_{mf}$. On account of the condensation of the $d$ bosons, it
turns out that the gauge fluctuations acquire a finite energy gap
through the Anderson-Higgs mechanism, so that the mean-field state
is stable against the gauge fluctuations. As a result, the physics
in the FM phase at energies much lower than the gap of gauge
bosons, $E_g$, can be described by the following effective
Hamiltonian in the grand canonical ensemble:
\begin{equation}
 H_{FM}=\sum_{\bm{k}}\left[(\epsilon_{\bm{k}}-\mu_f)f^{\dagger}_{\bm{k}}
       f_{\bm{k}}+\omega_{\bm{k}}b^{\dagger}_{\bm{k}}b_{\bm{k}}\right]
       , \label{qdefmh3}
\end{equation}
where $f_{\bm{i}}=\frac{1}{\sqrt{N}}\sum_{\bm{k}}e^{i\bm{k}\cdot
\bm{i}}f_{\bm{k}}$ and
$b_{\bm{i}}=\frac{1}{\sqrt{N}}\sum_{\bm{k}}e^{i\bm{k}\cdot
\bm{i}}b_{\bm{k}}$. (For the details of the derivation, see Sec.
\ref{gauge}.) The $f$ field describes the spinless charged
excitations
--- holons, with the dispersion relation
\begin{equation}
 \epsilon_{\bm{k}}= \! \left(1-\frac{x}{4}\right) \! t\sum_{\bm{u}}
         \cos{(\bm{k}\cdot\bm{u})} \ , \label{qdefmh1}
\end{equation}
and we shall see that the $b$ field describes the FM spin waves
--- magnons, with the dispersion relation
\begin{equation}
 \omega_{\bm{k}}=\frac{\eta t}{2}\left[3-\sum_{\bm{u}}
       \cos{(\bm{k}\cdot\bm{u})}\right] . \label{qdefmm1}
\end{equation}
In Eq. (\ref{qdefmh3}), we have neglected the interactions between
holons and magnons, which are irrelevant operators in the sense of
the renormalization group (RG). Equation (\ref{qdefmh3}) indicates
that there is spin-charge separation at low energy in the FM
metallic phase. (We will come back to this point in Sec.
\ref{gauge}.) Further, we shall see later that the bandwidth of
holons is of the order of $t$, while that of magnons is of the
order of $0.1t$.

\section{Results in the ferromagnetic metallic phase}
\label{result}
\subsection{Magnetic properties}
\paragraph{Magnon dispersion}

We shall first employ our mean-field theory to calculate the
transverse spin-spin correlation function, which is defined as
\begin{equation}
 i{\mathcal S}_{\perp}(t,\bm{x})\equiv \Theta (t)\langle
           [S^+_{\bm{i}}(t),S^-_{\bm{j}}(0)]\rangle \ , \label{tsscf1}
\end{equation}
where $S^{\pm}_{\bm{i}}=S^x_{\bm{i}}\pm iS^y_{\bm{i}}$ and
$\bm{x}=\bm{i}-\bm{j}$. In the large $J_H$ limit, $S_{\pm}$ can be
expressed by the $d$ bosons
\begin{equation}
 S^+_{\bm{i}}=d^{\dagger}_{\bm{i}\uparrow}d_{\bm{i}\downarrow} \ , ~~
 S^-_{\bm{i}}=d^{\dagger}_{i\downarrow}d_{i\uparrow} \ . \label{tsscf2}
\end{equation}
The Fourier transform of ${\mathcal S}_{\perp}(t,\bm{x})$, denoted
by ${\mathcal S}_{\perp}(\omega ,\bm{q})$, can be obtained from
the corresponding Matsubara function through analytical
continuation. In the FM phase, ${\mathcal S}_{\perp}(\omega
,\bm{q})$ can be related to the two-point correlation function of
the $b$ field within the mean-field theory:
\begin{equation}
 {\mathcal S}_{\perp}(\omega ,\bm{q})\approx x_d(T)S^T(\omega ,\bm{q})
       \ , \label{tsscf3}
\end{equation}
where $x_d(T)=\langle
d^{\dagger}_{\bm{i}\uparrow}d_{\bm{i}\uparrow}\rangle$ is the
average number of $d$ bosons in the condensate per site at
temperature $T$ and $S^T(\omega ,\bm{q})$ is the Fourier transform
of the retarded Green function of the $b$ field. In view of Eq.
(\ref{tsscf3}), one may identify the excitations corresponding to
the $b$ field as the magnons.

\begin{figure}
\begin{center}
 \includegraphics[width=1\columnwidth]{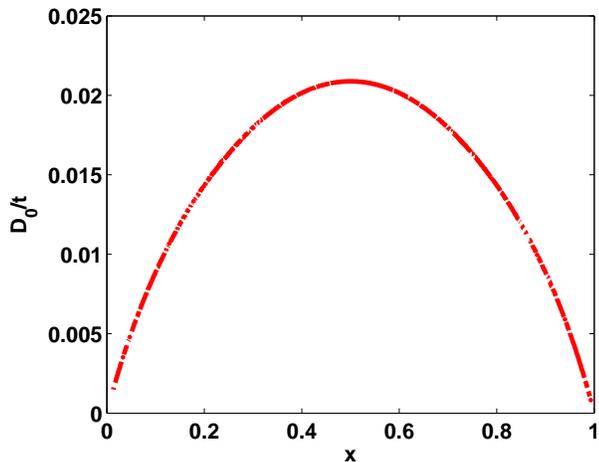}
 \caption{(Color online) The doping dependence of the spin stiffness at $T=0$.
         Here we set the lattice constant $l=1$.}
 \label{md0}
\end{center}
\end{figure}

We notice that the form of the magnon dispersion we obtained [Eq.
(\ref{qdefmm1})] is identical to that given by the FM Heisenberg
model on a simple cubic lattice with nearest-neighbor interactions
only. This result is similar to that predicted by previous studies
on the DE model.\cite{Zener, Anderson, deG,
spin-wave-zeroU,Furukawa} However, within the present framework, the
Heisenberg-type behavior for the magnon dispersion is a mean-field
result. By taking into account the so far ignored irrelevant (in the
sense of the RG) interactions between magnons, a deviation from the
Heisenberg spectrum is expected. On the general ground of the RG, we
expect that the deviation from the mean-field result (the Heisenberg
spectrum) will become more noticeable away from the zone center.
Furthermore, the small values of the spin stiffness at $T=0$ in the
regions $x\to 0$ and $x\to 1$ (see Fig. \ref{md0}) suggest that the
mean-field results may receive considerable corrections in these
regions. On the other hand, for the physically interested doping
regime $0.2<x<0.5$, the mean-field results should be robust. Recent
works on the DE model, such as the spin-wave theory based on the
large $S_c$ expansion\cite{Golosov2,Shannon} and an exact
calculation on a finite ring\cite{Kaplan} in the limit $J_H/t \gg
1$, have revealed a deviation of the magnon dispersion from the
Heisenberg spectrum in the DE model. The deviation is prominent at
large momenta,\cite{Golosov2,Kaplan} and the overall deviation is
very small in the doping range $0.2<x<0.6$.\cite{Kaplan} Our theory
is consistent with these results.

A Heisenberg-type behavior for the magnon dispersion may provide a
reasonably accurate picture for manganese oxides with large values
of $T_c$.\cite{magnon-dispersion} However, recent experiments
indicate deviations from this canonical behavior in compounds with
lower values of $T_c$. In particular, the softening of the mangon
dispersion near the boundary of the Brillouin zone is
observed.\cite{softening} Such an issue clearly depends on the
details of the short-distance physics, and is beyond the scope of
the present work. A systematic calculation which incorporates the
magnon-magnon and magnon-holon interactions, such as those given by
Eq. (\ref{fmeff2}), may be helpful.\cite{softening2} However, for a
complete comparison with experimental data, additional ingredients,
such as orbital fluctuations and orbital-lattice couplings, may also
need to be taken into account.\cite{Khaliullin2}

A finite damping rate of magnons at $T=0$ in the DE model has been
pointed out in Refs. \onlinecite{Golosov2}, \onlinecite{Shannon},
and \onlinecite{Kaplan2}. To address this problem within the
slave-fermion theory, we must go beyond the mean-field results. In
fact, the magnon acquires a finite lifetime, even at $T=0$, at the
two-loop order of the magnon self-energy diagrams by taking into
account the interactions between holons and magnons. [The leading
terms of these interactions are given by Eq. (\ref{fmeff2}).] The
physical origin of a nonvanishing $T=0$ damping rate of magnons can
be easily understood based on our spin-charge separated mean-field
state: Because the magnon dispersion, which is proportional to $k^2$
near the zone center, is definitely immersed into the
``particle-hole" continuum of holons in three dimensions, the magnon
can decay by exciting a single ``particle-hole" pair and another
magnon, which is similar to the physics of Landau damping in the
Fermi liquids.\cite{FL} Since the interactions between holons and
magnons are irrelevant operators in the sense of the RG, we expect
that the damping rate of the magnon near the zone center ($k=0$) is,
at least, proportional to $k^{\alpha}$ with $\alpha \geq 4$ in three
dimensions by the power counting usually employed in the
momentum-space RG. (Additional on-shell constraints may further
increase the value of $\alpha$.) The spin-wave theory in the
$J_H/t\rightarrow +\infty$ limit predicts that the damping rate of
the magnon near the zone center is proportional to $k^6$ in three
dimensions,\cite{Golosov2,Shannon} which implies that it is indeed
due to the irrelevant interactions. A full calculation by
incorporating these irrelevant operators is beyond the scope of this
paper, and their effects within the present framework are reserved
for further studies.

\paragraph{Doping dependence of the spin stiffness at $T=0$}

From Eq. (\ref{qdefmm1}), the magnon dispersion in the long
wavelength limit $\bm{k}\rightarrow 0$ is given by
$\omega_{\bm{k}}\approx D_0\bm{k}^2$ where
\begin{equation}
 D_0=\frac{\eta tl^2}{2S} \ , \label{qdefmm3}
\end{equation}
is the spin stiffness at $T=0$ and $l$ is the lattice constant.
Thus, the doping dependence of $D_0$ can be extracted from that of
$\eta$. The result is shown in Fig. \ref{md0}.

Our theory predicts that $D_0$ is symmetric in $x$ with respect to
the quarter-filling $x=0.5$. This result is similar to that
predicted by the spin-wave theory of the DE model in the limits
$J_H/t\rightarrow +\infty$ and $S_c\rightarrow
+\infty$,\cite{Golosov2} except that the magnitude of $D_0(x)$
around $x=0.5$ we obtained is smaller. The values of $D_0$ we
obtained are $D_0/(tl^2)=0.0144,0.018,0.0202$ for $x=0.2,0.3,0.4$.
Using the relation $D_0=JS_{eff}l^2$, where $S_{eff}=2-x/2$ is the
average spin for each site and $J$ denotes the effective exchange
coupling between spins, one may get $J=0.0076t,0.0097t,0.0112t$ for
$x=0.2,0.3,0.4$, which correspond to $J=2.28,2.91,3.36$ meV if we
use $t=0.3$ eV. $J\approx 1.9, 2.4$ meV found in
La$_{0.8}$Sr$_{0.2}$MnO$_3$\cite{magnon-dispersion} and
La$_{0.7}$Pb$_{0.3}$MnO$_3$,\cite{Woodfield} respectively. We see
that the values of $J$ we obtained are consistent with those
extracted from experimental data. To sum up, for the magnon
dispersion relation, the difference between our mean-field theory
and the spin-wave theory of the DE model in the doping range
$0.2<x<0.5$ mainly lies at the doping dependence of the spin
stiffness at $T=0$ and its magnitude.

Finally, we mention that the low temperature magnetization in the FM
phase can be easily calculated within the present mean-field theory,
and it is
\begin{equation}
 M(T)=M_0\left[1-\frac{\zeta (3/2)}{M_0}\left(\frac{T}{4\pi D_0}
     \right)^{3/2}+\cdots \right] , \label{m3}
\end{equation}
which is identical to that predicted by the simple cubic Heisenberg
ferromagnet, where $M_0=(2-x/2)l^{-3}$ is the magnetization at
$T=0$.

\subsection{Estimation of $T_c$}

\begin{figure}
\begin{center}
 \includegraphics[width=1\columnwidth]{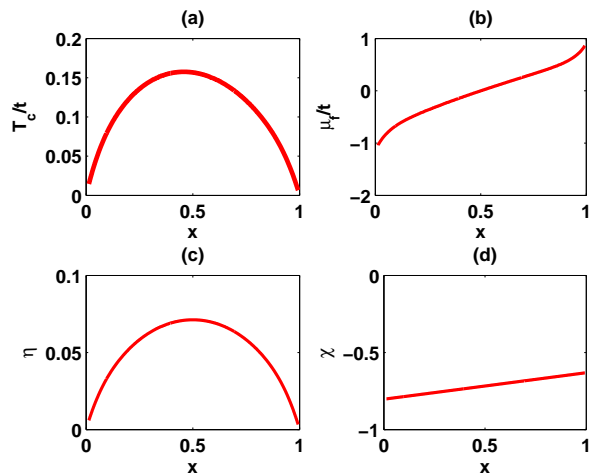}
 \caption{(Color online) (a) $T_c/t$ v.s. $x$; (b) $\mu_f/t$ v.s. $x$ at $T=T_c$; (c) $\eta$
         v.s. $x$ at $T=T_c$; (d) $\chi$ v.s. $x$ at $T=T_c$.}
 \label{mntc}
\end{center}
\end{figure}

Next, we would like to estimate $T_c$ within our mean-field theory,
which can be obtained from the mean-field equations [Eqs.
(\ref{qdemfe1}) --- (\ref{qdehn1})] numerically by setting $\Delta
=\Delta_0$. The numerical results are shown in Fig. \ref{mntc}. A
few points about our results should be discussed. First of all,
$T_c$ is asymmetric in $x$ with respect to the point $x=0.5$ though
$\eta (T_c)$ is still symmetric in $x$ with respect to $x=0.5$. This
is in contrast to $D_0$. This asymmetry suggests that the FM phase
is more robust at $x<0.5$ by including strong on-site Coulomb
repulsions and Hund's rule couplings properly. Next, the previous
work based on the dynamical mean-field theory gives an estimate of
$T_c$ for $J_H/t\gg 1$: $T_c/t=0.146$ at $x=0.3$.\cite{Furukawa} Our
mean-field theory predicts $T_c/t=0.1454$ at $x=0.3$, which is quite
close to the value obtained from the dynamical mean-field theory.
Finally, the maximum of $T_c/t$ is reached at $x=0.458$ with the
value $T_c/t=0.1575$, which corresponds to $T_c\sim 548$ K for
$t\sim 0.3$ eV. This value is about $2$ times of that obtained by
experiments. As usual, the fluctuations will reduce the mean-field
value. Nevertheless, the difference between the mean-field result
and experimental data may still not be explained even if we include
the fluctuations. This may not be surprising since, after all, $T_c$
is a non-universal quantity and the above result simply indicates
that to have a good estimation of $T_c$ a few ingredients that are
ignored in the QDE model, such as the AF superexchange interactions
between core spins, orbital fluctuations, and electron-phonon
interactions, have to be included. In fact, experiments on the
oxygen-isotope substitution show that phonons are important in the
determination of $T_c$.\cite{phonon}

\subsection{The low-temperature specific heat}

Low-temperature ($T\leq 10$ K) heat-capacity measurements provide
information regarding the bulk properties of solids. For a
magnetic solid, the low-temperature specific heat is composed of
numerous contributions and it is typically given by
\begin{equation}
 c_v=c_{elec}+c_{hyp}+c_{lat}+c_{mag} \ . \label{mnsh1}
\end{equation}
Here $c_{elec}$ is the electronic contribution, which takes the form
$c_{elec}=\gamma T$, $c_{hyp}$ arises from the hyperfine field of
the nuclear moment, $c_{lat}$ is the contribution of phonons, and
$c_{mag}$ is the contribution from the magnetic spin waves, which is
usually estimated as the form $\sum_nB_nT^n$.

In our theory, $c_{elec}$ is primarily given by the holon sector due
to the spin-charge separation at low energy, yielding
\begin{eqnarray*}
 \gamma =\frac{(\pi k_B)^2}{3}~g(\epsilon_F) \ ,
\end{eqnarray*}
away from the van Hove singularities, where $k_B$ is the Boltzmann
constant, and $g(\epsilon_F)$ is the density of states (DOS) of
holons at the Fermi energy.

\begin{figure}
\begin{center}
 \includegraphics[width=1\columnwidth]{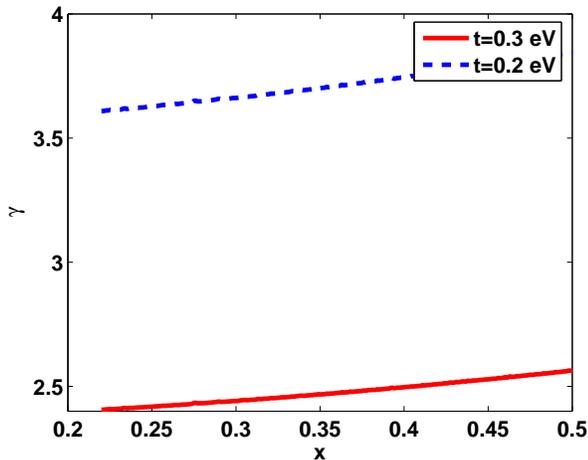}
 \caption{(Color online) $\gamma =c_{elec}/T$, in unit of mJ/mol K$^2$, in the doping
         range $0.2<x<0.5$.}
 \label{mncv}
\end{center}
\end{figure}

Figure \ref{mncv} shows the values of $\gamma$ in the doping range
$0.2<x<0.5$ with $t=0.3$ eV (solid line) and $t=0.2$ eV (dashed
line). The magnitudes of $\gamma$ we obtained are consistent with
the experimental data. Furthermore, the mean-field theory predicts a
very weak doping dependence of $\gamma$, which is approximately
given by
\begin{equation}
 \gamma \propto \frac{1}{1-x/4} \ . \label{mncv1}
\end{equation}
Such a weak doping dependence of $\gamma$ may not contradict to
the experimental data at $x=0.2$ and $0.3$, where the measured
values of $\gamma$ ($\approx 3.3$ mJ/mol K$^2$) appear to be
nearly a constant in the FM metallic phase.\cite{Woodfield} More
detailed experiments is warranted to resolve this issue.

\begin{figure}
\begin{center}
 \includegraphics[width=1\columnwidth]{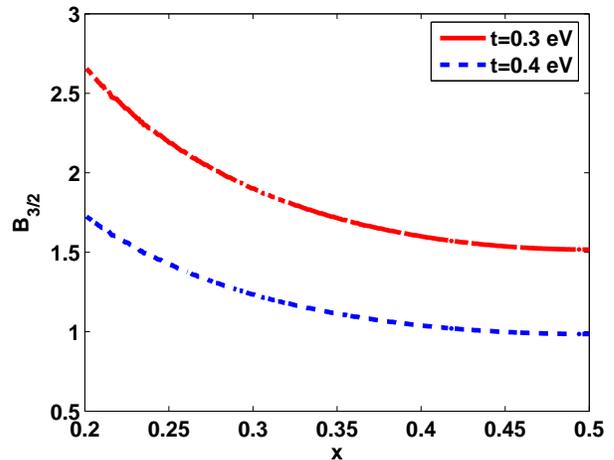}
 \caption{(Color online) $B_{3/2}=c_{mag}/T^{3/2}$, in unit of mJ/mol K$^{5/2}$, in
         the doping range $0.2<x<0.5$.}
 \label{cvb32}
\end{center}
\end{figure}

As for $c_{mag}$, the magnon contribution to the low-temperature
specific heat is similar to that in the simple cubic Heisenberg
ferromagnet and takes the form
\begin{eqnarray*}
 c_{mag}=B_{3/2}T^{3/2} \ ,
\end{eqnarray*}
with
\begin{eqnarray*}
 B_{3/2}=\frac{15\zeta (5/2)k_B^{5/2}}{32(\pi D_0)^{3/2}} \ .
\end{eqnarray*}
The values of $B_{3/2}$ in the doping range $0.2<x<0.5$ are shown in
Fig. \ref{cvb32} with $t=0.3$ eV (solid line) and $t=0.4$ eV (dotted
line). The values of $B_{3/2}$ we obtained are close to the
experimental data\cite{Woodfield} ($B_{3/2}=1.1802$
mJ$/$K$^{5/2}$~mol for $x=0.2$ and $B_{3/2}=0.95672$
mJ$/$K$^{5/2}$~mol for $x=0.3$) if we use $t=0.4$ eV.

To sum up, the consistency of the magnitudes of $\gamma$ and
$B_{3/2}$ in the doping range $0.2<x<0.5$ evaluated in terms of the
mean-field theory suggests that the low-temperature thermodynamics
of the FM metallic phase can be well-described by the one-orbital
QDE model, and this result can be viewed as an indirect evidence of spin-charge
separation at low energy in the FM metallic phase.

\subsection{Optical conductivity at low energy}

\begin{figure}
\begin{center}
 \includegraphics[width=1\columnwidth]{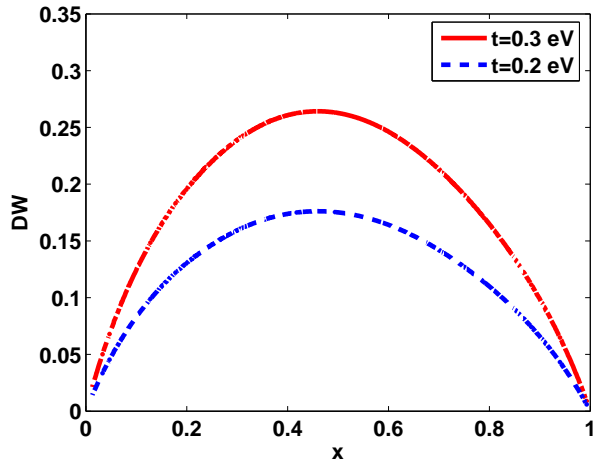}
 \caption{(Color online) $DW$ as a function of $x$.}
 \label{mndr}
\end{center}
\end{figure}

One of the consequences of the spin-charge separation in our
theory is that the optical conductivity at low frequency ($\omega
\ll E_g$) is mainly contributed by free holons, where $E_g$
denotes the gap of gauge fluctuations. Especially, the Drude
weight is completely determined by free holons, which is given by
\begin{equation}
 {\mathcal D}=-\frac{\pi e^2}{2l}~K \ , \label{mnoc2}
\end{equation}
where $K$ is the average kinetic energy of holons per site and
$l\approx 3.9$ \AA ~is the lattice constant. Following experiments,
one may measure the Drude weight in unit of $\pi e^2/(2m_el^3)$,
denoted by $DW$, where $m_e$ is the electron mass. Within our
mean-field theory, $DW$ at $T=0$ is
\begin{equation}
 DW=\frac{6m_etl^2\eta}{\hbar^2}\left(1-\frac{x}{4}\right) , \label{mnoc21}
\end{equation}
where $\hbar$ is the Planck constant. The doping dependence for
$DW$, given by Eq. (\ref{mnoc21}), is shown in Fig. \ref{mndr}
with $t=0.3$ eV (solid line) and $t=0.2$ eV (dotted line). From
it, we see that, in the doping range $0.2<x<0.5$, $DW\sim 0.19 -
0.26$ for $t=0.3$ eV, and $DW\sim 0.13 - 0.18$ for $t=0.2$ eV.
Since the Drude peak at $T=0$ does not carry $100\%$ of the
weight, this result implies that some spectral weight must be
transferred to higher energies ($\omega \geq E_g$) due to the
optical conductivity sum rule, so that the Drude weight is
suppressed compared with normal metals.

Optical conductivity spectra have been investigated for single
crystals of La$_{1-x}$Sr$_x$MnO$_3$ with $0\leq x\leq
0.3$.\cite{Okimoto} The peculiar behaviors observed in the
low-energy optical spectra ($\omega <0.1$ eV) in the FM metallic
phase at low temperature, which cannot be explained by the simple
Fermi-liquid picture, are as follows: (i) The low-energy spectra
are composed mostly of the incoherent part and lightly of the
Drude response (about $20\%$ - $30\%$ in fraction). (ii) The Drude
part is discernible below $0.04$ eV, but with an anomalously small
spectral weight. For example, the value of $DW$ is as small as
$0.012$ even for the lowest temperature spectra for both $x=0.175$
and $0.3$. Further optical studies on
La$_{0.7}$Ca$_{0.3}$MnO$_3$\cite{Kim} confirm that, in the FM
metallic phase at very low temperatures, the low-energy optical
conductivity spectra ($\omega <0.5$ eV) show two types of
absorption features: a sharp Drude peak with little weight (about
$33\%$ in fraction or $DW\sim 0.02$) superimposed to the broad
incoherent absorption band. Our analysis indicates that in the FM
metallic phase the optical conductivity spectra of the QDE model
at low energy ($\omega \ll J_H$) indeed consists of two parts:
\begin{eqnarray*}
 \sigma_1(\omega)\sim \sigma_c(\omega)+\sigma_{inc}(\omega) \ ,
\end{eqnarray*}
where $\sigma_c(\omega)$ is the coherent (Drude) part, which is
mainly contributed by holons, and $\sigma_{inc}(\omega)$
represents the incoherent part with most of its weight in the
region $\omega \geq E_g$, which arises from the strong scattering
between gauge fields and holons. Thus, according to our theory,
the energy scale below which the Drude part is discernible can be
regarded as the lower bound of $E_g$. A rough estimate of the
value of $E_g$ ($E_g\sim 0.038-0.071$ eV in the doping range
$0.15<x<0.5$ for $t\sim 0.3$ eV) is consistent with the
experiment. (See Sec. \ref{gauge}.) This picture is consistent
with the observed optical spectra at low energy. However, the
values of $DW$ we obtained are larger than the experimental data
by an order of magnitude. Since our mean-field theory is supposed
to be accurate at low energy, the quantitative discrepancy between
our results and the experimental data indicates that to explain
the optical spectra of manganites, other ingredients beyond the
one-orbital QDE model must be taken into account. Some possible
scenarios, such as orbital fluctuations\cite{Horsch} or
electron-phonon conplings,\cite{Millis,Edwards} have been
proposed. Applying our method to these ``extended" QDE models is
an interesting problem.

\subsection{The electron spectral function and tunneling density of states}

The most important result in our mean-field theory is the
spin-charge separation at low energy. A direct examination of this
phenomenon is to study the behavior of the electron spectral
function at low temperatures, which can be measured by the ARPES
experiments. The electron spectral function is defined as
\begin{eqnarray}
 & & A(\omega ,\bm{q})\equiv -2{\mathrm Im}\{{\mathcal G}
     (i\omega_n\rightarrow \omega ,\bm{k})\} \ . \label{qdeesf1}
\end{eqnarray}
Here ${\mathrm Im}\{\cdots\}$ means the imaginary part of $\cdots$,
and ${\mathcal G}(K)$ with $K=(i\omega_n,\bm{k})$ is the Fourier
transform of the electron Green function, which is defined by
\begin{eqnarray}
 {\mathcal G}(X) &\equiv& -\langle {\mathcal T}\{\bar{c}_{\bm{i}\sigma}
          (\tau)\bar{c}^{\dagger}_{\bm{j}\sigma}(0)\}\rangle \nonumber
          \\
          &=& -\frac{1}{2S}\langle {\mathcal T}\{f_{\bm{i}}^{\dagger}(\tau)
          d_{\bm{i}\sigma}(\tau)d^{\dagger}_{\bm{j}\sigma}(0)f_{\bm{j}}
          (0)\}\rangle \ , \label{qdeesf11}
\end{eqnarray}
where $X=(\tau ,\bm{i}-\bm{j})$.

At the mean-field level, holons and magnons are decoupled from one
another and $d_{\bm{i}\uparrow}$ is treated as a $c$-number, in
stead of an operator. As a result, the electron Green function can
be approximated as
\begin{eqnarray}
 {\mathcal G}(X) &\approx& \frac{1}{2S}\langle {\mathcal T}\{f_{\bm{j}}
          (0)f_{\bm{i}}^{\dagger}(\tau)\}\rangle \langle
          d^{\dagger}_{\bm{i}\uparrow}d_{\bm{i}\uparrow}\rangle \nonumber
          \\
          & & +\frac{1}{2S}\langle {\mathcal T}\{f_{\bm{j}}
          (0)f_{\bm{i}}^{\dagger}(\tau)\}\rangle \langle {\mathcal T}\{
          d_{\bm{i}\downarrow}(\tau)d^{\dagger}_{\bm{j}\downarrow}(0)\}
          \rangle \nonumber \\
          &=& -\frac{x_d(T)}{2S}~G(-X)+\frac{1}{2S}~G(-X)S(X) \ ,
          \label{qdeesf12}
\end{eqnarray}
where $x_d(T)$ is determined by the equation
\begin{equation}
 x_d(T)+\frac{1}{N}\sum_{\bm{k}}n_B(\omega_{\bm{k}})=2S-x \ , \label{schn1}
\end{equation}
and
\begin{eqnarray*}
 G(X) &=& -\langle {\mathcal T}\{f_{\bm{i}}(\tau)f_{\bm{j}}^{\dagger}(0)\}
      \rangle \ , \\
 S(X) &=& -\langle {\mathcal T}\{b_{\bm{i}}(\tau)b^{\dagger}_{\bm{j}}(0)\}
      \rangle\ ,
\end{eqnarray*}
are the propagators of holons and magnons, respectively. In terms of
Eq. (\ref{qdeesf12}), the electron spectral function is given
by\cite{sum-rule}
\begin{equation}
 A(\omega ,\bm{k})=Z(x,T)~\delta (\omega +\epsilon_{\bm{k}}-\mu_f)
          +a(\omega ,\bm{k}) \ . \label{qdeesf2}
\end{equation}
where $Z(x,T)=\pi x_d(T)/S$ and
\begin{eqnarray*}
 a(\omega ,\bm{k}) &=& \frac{\pi}{SN}\sum_{\bm{q}}[n_B(\omega_{\bm{k}+\bm{q}})
          +n_F(\epsilon_{\bm{q}}-\mu_f)] \\
          & & \times ~\delta (\omega +\epsilon_{\bm{q}}-\mu_f
          -\omega_{\bm{k}+\bm{q}}) \ .
\end{eqnarray*}

\begin{figure}
\begin{center}
 \includegraphics[width=1\columnwidth]{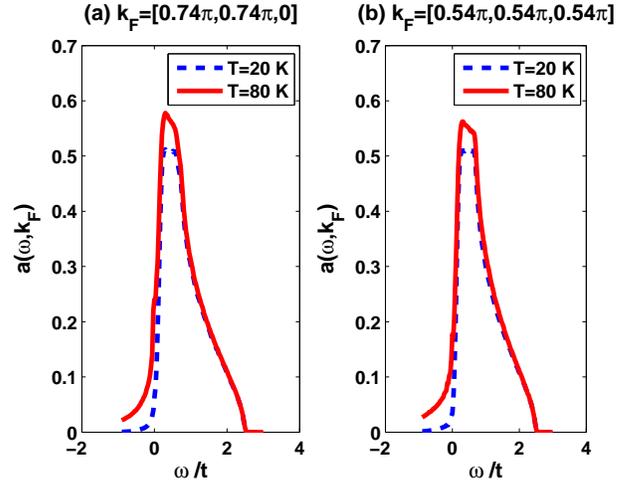}
 \caption{(Color online) The electron spectral function, with the subtraction of the
         quasiparticle peak, at the hole concentration $x=0.4$.
         $\bm{k}_F$ denotes the Fermi momentum. The temperatures we
         consider are $T/t=1/174$ (dashed line) and $25/1088$ (solid line),
         which correspond to $T=20, 80$ K, respectively, for $t=0.3$
         eV.}
 \label{mnesf1}
\end{center}
\end{figure}

\begin{figure}
\begin{center}
 \includegraphics[width=1\columnwidth]{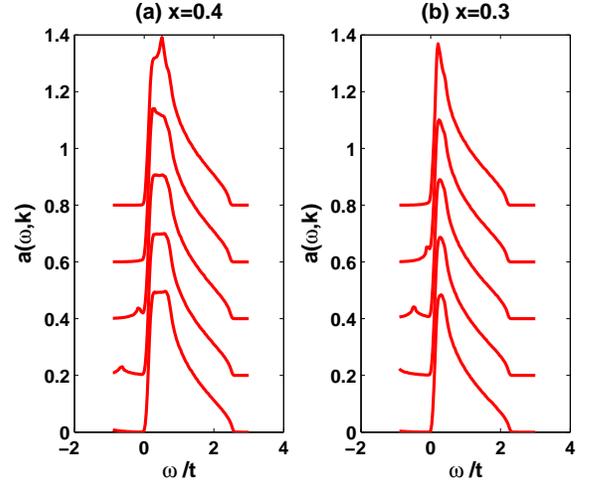}
 \caption{(Color online) The electron spectral function, with the subtraction of the
         quasiparticle peak, at the temperature $T/t=1/174$ and hole
         concentration (a) $x=0.4$ and (b) $x=0.3$. Here $\bm{k}=k_0
         \times [\pi,\pi,0]$ with $k_0=0.5, 0.6, 0.7, 0.8, 0.9$ from bottom to top.}
 \label{mnesf2}
\end{center}
\end{figure}

Equation (\ref{qdeesf2}) indicates that the electron spectral
function at low energy consists of two parts: a sharp quasiparticle
peak, following the holon dispersion, superimposed to a broad
incoherent part given by $a(\omega ,\bm{k})$.  The spectral weight
of the quasiparticle peak is given by $Z(x,T)$, which is associated
with the condensate of the $d$ bosons. At $T\ll T_c$, it is given by
\begin{equation}
 Z(x,T)\approx 2\pi \! \left[1-\frac{x}{4}-\frac{\zeta (3/2)}{4} \!
       \left(\frac{Tl^2}{4\pi D_0}\right)^{3/2}\right] . \label{epw1}
\end{equation}
By taking into account impurity scattering and the interactions
between holons and magnons, the $\delta$-function peak will be
broadened, and the width $1/\tau$ is given by $1/\tau
=max\{1/\tau_0,k_BT\}$ where $\tau_0$ is the lifetime due to
impurity scattering. The low-temperature behavior of $a(\omega
,\bm{k}_F)$ at the hole concentration $x=0.4$ is shown in Fig.
\ref{mnesf1}, where $\bm{k}_F\approx 0.74\pi (1,1,0)$ and $0.54\pi
(1,1,1)$ in Fig. \ref{mnesf1} (a) and (b), respectively. Figure
\ref{mnesf2} shows $a(\omega ,\bm{k})$ along the direction from
$[0,0,0]$ to $[\pi ,\pi ,0]$, at the temperature $T/t=1/174$ and the
hole concentration (a) $x=0.4$ and (b) $x=0.3$. A common feature of
$a(\omega ,\bm{k})$ is that it exhibits a broad asymmetric peak at
low temperature with the width $\sim t$, and the peak position is at
$\omega \sim 0.5t$ and $0.21t$ for $x=0.4$ and $0.3$, respectively.
Furthermore, at low temperature, the peak position and its width are
insensitive to the variations of temperatures. This asymmetric peak
implies the violation of the particle-hole symmetry. As raising the
temperature, we can see a transfer of the spectral weight from the
quasiparticle peak to the incoherent part. Such a trend persists
until at the critical temperature where the coherent part disappears
completely, i.e. $Z(x,T_c)=0$. The presence of such a broad
asymmetric peak at low temperature reflects the composite nature of
electrons at low energy in the large $U, J_H$ limit, i.e. a
manifestation of the spin-charge separation at low energy, which can
be observed in the photoemission experiment.

A similar form for the electron spectral function [Eq.
(\ref{qdeesf2})] was also obtained by a mean-field approximation to
the DE model in the limit $J_H/t\rightarrow +\infty$.\cite{Sarkar}
There, the FM phase also corresponds to the condensed phase of the
Schwinger bosons. However, the doping dependence of the mean-field
parameters in that approach is different from ours, especially the
bandwidth of holons. This results in distinct doping dependence for
various physical quantities, such as the spin stiffness at $T=0$ and
the Drude weight. Moreover, quadratic approximations for the
dispersion relations of holons and magnons were used to calculate
the electron spectral function, which leads to a weak logarithmic
singularity in the incoherent part $a(\omega ,\bm{k})$ with the
width being of the order of the maximum magnon energy. We do not see
such a singularity, as can be seen in Figs. \ref{mnesf1} and
\ref{mnesf2}. Further, the width of the broad peak we obtained is
much larger, which is consistent with the ARPES measurements.

\begin{figure}
\begin{center}
 \includegraphics[width=1\columnwidth]{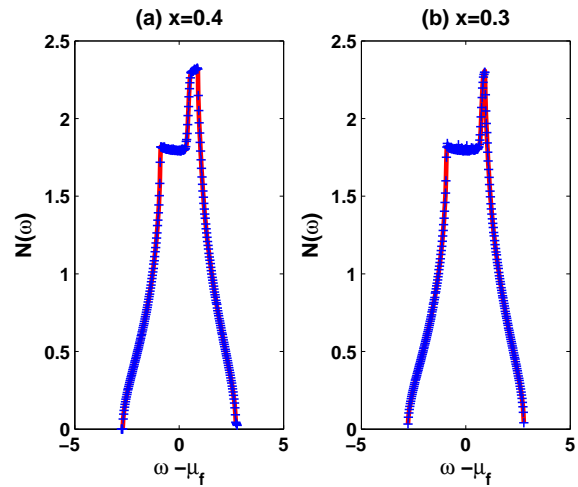}
 \caption{(Color online) The tunneling DOS at the temperatures $T/t=1/174$ (solid line)
         and $25/1088$ ($+$) with the hole concentration (a) $x=0.4$
         and (b) $x=0.3$. We have set $t=1$.}
 \label{mntdos1}
\end{center}
\end{figure}

Another quantity which is related to $A(\omega ,\bm{k})$ is the
tunneling DOS:
\begin{equation}
 N(\omega)=\frac{1}{N}\sum_{\bm{k}}A(\omega ,\bm{k}) \,
  \label{tdos1}
\end{equation}
which can be measured by the scanning tunneling spectroscopy
(STS). Within the mean-field theory, $N(\omega)$ can be written as
\begin{equation}
 N(\omega)=\frac{Z(x,T)}{t(1-x/4)}~N_3 \! \left(\frac{\bar{\omega}}
  {1-x/4}\right) \! +\frac{1}{N}\sum_{\bm{k}}a(\omega ,\bm{k}) \ ,
  \label{tdos3}
\end{equation}
where $N_3(\epsilon)\equiv \frac{1}{N}\sum_{\bm{k}}\delta \!
\left[\epsilon -\sum_{\bm{u}}\cos{(\bm{k}\cdot\bm{u})}\right]$ and
$\bar{\omega}=(\omega -\mu_f)/t$. Figure \ref{mntdos1} shows
$N(\omega)$ as a function of $\bar{\omega}$ at two temperatures
$T/t=1/174$ and $25/1088$ with the hole concentration (a) $x=0.4$
and (b) $x=0.3$. A few salient features about the low-temperature
behavior of the tunneling DOS are as follows: (i) $N(\omega)$
consists of two parts: One is proportional to the DOS for a
tight-binding model on a simple cubic lattice, which is contributed
by the quasiparticle peak of the electron spectral function, and the
other exhibits an asymmetric peak, which results from the incoherent
part of the electron spectral function. (ii) The position of the
asymmetric peak is at $\omega -\mu_f\sim \chi (T=0)t$, the
renormalized hopping amplitude of holons. (iii) The values of
$N(\omega)$ is insensitive to the variation of temperatures. Again,
this structure of $N(\omega)$ is a natural consequence of
spin-charge separation.

Variable temperature STS studies on single crystals of
La$_{0.7}$Pb$_{0,3}$MnO$_3$ have been done in the temperature
range $100 - 375$ K.\cite{TDS-STS} Experimental data show that
$N(\epsilon_F)$ grows rapidly below $T_c$ and reaches a
temperature independent value, where $\epsilon_F$ is the Fermi
energy. Furthermore, $N(\omega)$ near $\epsilon_F$ is flat in the
scale of $T$ below $T_c$. Further STS studies on single crystals
of La$_{0.7}$Ca$_{0,3}$MnO$_3$\cite{TDS-STS2} confirm these
results. Our prediction about the behavior of $N(\omega)$ near the
Fermi energy is consistent with experimental results. Further STS
studies at low temperatures must be performed to verify our
mean-field theory. Especially, an observation of the asymmetric
peak away from the Fermi energy in the tunneling DOS at low
temperatures will be solid confirmation of the slave-fermion
theory.

\section{Gauge fluctuations and the stability of the mean-field state}
\label{gauge}

In order to study the role of gauge fluctuations and to derive the
low energy effective theory, we can start with Eq. (\ref{qdell1}),
and parameterize the Hubbard-Stratonovich fields as
\begin{equation}
 \chi_{\bm{i}+\bm{u},\bm{i}}=\chi e^{-i\phi_{\bm{i},\bm{u}}} \ , ~~
 \eta_{\bm{i}+\bm{u},\bm{i}}=\eta e^{-i\tilde{\phi}_{\bm{i},\bm{u}}}
     \ . \label{urvb3}
\end{equation}
One may further decompose $\phi_{\bm{i},\bm{u}}$ and
$\tilde{\phi}_{\bm{i},\bm{u}}$ by
\begin{equation}
 \phi_{\bm{i},\bm{u}}={\mathcal A}_{\bm{i},\bm{u}}+{\mathcal B}_{\bm{i},\bm{u}}
     \ , ~~
 \tilde{\phi}_{\bm{i},\bm{u}}={\mathcal A}_{\bm{i},\bm{u}}-
     {\mathcal B}_{\bm{i},\bm{u}} \ . \label{u1phase1}
\end{equation}
Such a decomposition becomes transparent once we realize that under
the U($1$) gauge transformation, ${\mathcal A}_{\bm{i},\bm{u}}$
transforms like a gauge field, while ${\mathcal B}_{\bm{i},\bm{u}}$
is gauge invariant. Simple manipulations show that the ${\mathcal
B}_{\bm{i},\bm{u}}$ field acquires a finite energy gap. Furthermore,
the amplitude fluctuations of $\chi_{\bm{i}+\bm{u},\bm{i}}$ and
$\eta_{\bm{i}+\bm{u},\bm{i}}$ are also gapped. For the energy below
these mass gaps, one may integrate out the massive fields and take
the continuum limit
\begin{eqnarray}
 & & f_{\bm{i}}\rightarrow l^{d/2}h(X) \ , ~~
     d_{\bm{i}\sigma}\rightarrow l^{d/2}d_{\sigma}(X) \ , \nonumber
     \\
 & & {\mathcal A}_{\bm{i},\bm{u}}\rightarrow \bm{u}\cdot
     \bm{a}(X) \ , ~~
     \lambda_{\bm{i}}\rightarrow \lambda (X) \ , \label{eff2}
\end{eqnarray}
where $X=(\tau ,\bm{x})$. After doing so, we arrive at the following continuum
Lagrangian within the effective-mass approximation
\begin{eqnarray}
 {\mathcal L} &=& \sum_{\sigma}d^{\dagger}_{\sigma}(\partial_{\tau}+i\lambda)
              d_{\sigma}+h^{\dagger}(\partial_{\tau}+i\lambda -\mu_0)h
              -2iM\lambda \nonumber \\
              & & +\frac{1}{2m_f}|(\bm{\nabla}-i\bm{a})h|^2+ \! \sum_{\sigma}
              \frac{1}{2m_b}|(\bm{\nabla}-i\bm{a})d_{\sigma}|^2 ,  ~~
              \label{deeffl1}
\end{eqnarray}
where $M=Sl^{-3}$. In view of Eq. (\ref{deeffl1}), the Lagrangian
multiplier $\lambda$ plays the role of the time component of the
gauge fields.

In the FM phase, without loss of generality, one may parameterize
$d_{\sigma}$ in the following way:
\begin{eqnarray}
 d_{\uparrow}=\sqrt{n_d+\rho}~e^{i\phi} \ , ~~ d_{\downarrow}=b
   \ , \label{leeff2}
\end{eqnarray}
on account of the BEC of the Schwinger bosons, where
$n_d=(2S-x)l^{-3}$ is the average density of $d$ bosons, and $\rho$
and $\phi$ describe the amplitude and phase fluctuations of
$d_{\uparrow}$, respectively. That is, we choose the direction of
the order parameter (magnetization) to be the $z$-axis. Inserting
Eq. (\ref{leeff2}) into Eq. (\ref{deeffl1}) gives rise to
\begin{eqnarray*}
 {\mathcal L}_{FM} &=& b^{\dagger}(\partial_{\tau}+i\lambda)b
            +h^{\dagger}(\partial_{\tau}+i\lambda -\mu_0)h-in
            \lambda \\
            & & +\frac{1}{2m_f}|(\bm{\nabla}-i\bm{a})h|^2+\frac{1}
            {2m_b}|(\bm{\nabla}-i\bm{a})b|^2 \\
            & & +\frac{1}{8m_bn_d}|\bm{\nabla}\rho|^2+\frac{n_d}
            {2m_b}|\bm{\nabla}\phi -\bm{a}|^2 \\
            & & +\frac{1}{2m_b}\rho |\bm{\nabla}\phi -\bm{a}|^2
            +i\rho (\lambda +\partial_{\tau}\phi) \ .
\end{eqnarray*}
The U(1) gauge structure of ${\mathcal L}_{FM}$ now reads
\begin{eqnarray}
 & & h\rightarrow he^{-iw} \ , ~~ b\rightarrow be^{-iw} \ ,
     \nonumber \\
 & & \rho \rightarrow \rho \ , ~~ \phi \rightarrow \phi -w \ ,
     \nonumber \\
 & & \bm{a}\rightarrow \bm{a}-\bm{\nabla}w \ , ~~ \lambda
     \rightarrow \lambda +\partial_{\tau}w \ . \label{u1gauge4}
\end{eqnarray}
With an eye on the gauge invariance of ${\mathcal L}_{FM}$, one may
choose the gauge $w=\phi$. This amounts to redefining the $h$,
$b$, $\bm{a}$ and $\lambda$ as follows:
\begin{eqnarray}
 & & \tilde{h}=he^{-i\phi} \ , ~~ \tilde{b}=be^{-i\phi} \ , \nonumber
     \\
 & & \tilde{\bm{a}}=\bm{a}-\bm{\nabla}\phi \ , ~~
     \tilde{\lambda}=\lambda +\partial_{\tau}\phi \ .
     \label{ugauge1}
\end{eqnarray}
We notice that $\tilde{h}$, $\tilde{b}$, $\tilde{\bm{a}}$, and
$\tilde{\lambda}$ are all gauge invariant. In terms of Eq.
(\ref{ugauge1}), ${\mathcal L}_{FM}$ becomes
\begin{eqnarray*}
 {\mathcal L}_{FM} &=& \tilde{b}^{\dagger}(\partial_{\tau}
            +i\tilde{\lambda})\tilde{b}+\tilde{h}^{\dagger}
            (\partial_{\tau}+i\tilde{\lambda}-\mu_0)\tilde{h}
            +i(\rho -n)\tilde{\lambda} \\
            & & +\frac{1}{2m_f}|(\bm{\nabla}-i\tilde{\bm{a}})
            \tilde{h}|^2+\frac{1}{2m_b}|(\bm{\nabla}
            -i\tilde{\bm{a}})\tilde{b}|^2 \\
            & & +\frac{1}{8m_bn_d}|\bm{\nabla}\rho|^2+\frac{1}
            {2m_b}(n_d+\rho)\tilde{\bm{a}}^2 .
\end{eqnarray*}
In the above, the higher order terms in $\rho$, i.e. the
self-interactions of $\rho$, have been neglected. This is because
they only generate irrelevant operators. Integrating out
$\tilde{\lambda}$ gives rise to the constraint
\begin{equation}
 \tilde{b}^{\dagger}\tilde{b}+\tilde{h}^{\dagger}\tilde{h}+\rho =n
       \ , \label{ndo5}
\end{equation}
which is nothing but the continuum version of the NDO condition
[Eq. (\ref{ndo3})]. Using Eq. (\ref{ndo5}), one may further
integrate out the $\rho$ field, yielding
\begin{eqnarray*}
 {\mathcal L}_{FM} &=& \tilde{h}^{\dagger}(\partial_{\tau}-\mu_0)
            \tilde{h}+\frac{1}{2m_f}|\bm{\nabla}\tilde{h}|^2
            +\tilde{b}^{\dagger}\partial_{\tau}\tilde{b}+D_0
            |\bm{\nabla}\tilde{b}|^2 \nonumber \\
            & & +\frac{1}{2}~m_a\tilde{\bm{a}}^2+\cdots \ ,
\end{eqnarray*}
where $\cdots$ represents the interactions between $\tilde{h}$,
$\tilde{b}$, and $\tilde{\bm{a}}$, $D_0=(2m_b)^{-1}$ is the FM
spin stiffness at $T=0$, and $m_a=[2M-(1-m_b/m_f)n]/m_b$.

We see that in the FM phase the gauge fields acquire a ``mass" term
through the Anderson-Higgs mechanism. In other words, the
excitations corresponding to $\tilde{\bm{a}}$ acquire a finite
energy gap $E_g\sim m_a/k_F$, up to a multiplicative constant of
order one, where $k_F$ is the Fermi momentum of holons. Using the
mean-field parameters, $m_f$ can be estimated as
$m_f^{-1}=(1-x/4)tl^2$, and we have
\begin{equation}
 \frac{E_g}{t}\sim \frac{4\sqrt{2}D_0/(tl^2)}{\sqrt{3+W/f(x)}}
      \left\{1-\frac{x}{4}\left[1-f(x) \! \left(\frac{tl^2}{2D_0}
      \right)\right]\right\} , \label{am1}
\end{equation}
where $W=\epsilon_F/t$ and $f(x)=1-x/4$. Equation (\ref{am1}) gives
rise to the values $E_g/t=0.1273 - 0.2381$ in the doping range
$0.15<x<0.5$. For $t\sim 0.3$ eV, this corresponds to $E_g\sim
0.038-0.071$ eV, which is about the same order of $T_c$.

When the energy is much lower than $E_g$, one may further
integrate out $\tilde{\bm{a}}$, the resulting effective Lagrangian
can be written as ${\mathcal L}_{FM}={\mathcal L}_0+{\mathcal
L}_{int}$ where
\begin{equation}
 {\mathcal L}_0=\tilde{h}^{\dagger}(\partial_{\tau}-\mu_f)
            \tilde{h}+\frac{1}{2m_f}|\bm{\nabla}\tilde{h}|^2
            +\tilde{b}^{\dagger}\partial_{\tau}\tilde{b}+D_0
            |\bm{\nabla}\tilde{b}|^2 , \label{fmeff1}
\end{equation}
and
\begin{equation}
 {\mathcal L}_{int}=\frac{g_1}{2}(\bm{j}_h+\bm{j}_b)^2+\frac{g_2}
           {2}(\bm{\nabla}\rho_h+\bm{\nabla}\rho_b)^2 \ .
           \label{fmeff2}
\end{equation}
In Eq. (\ref{fmeff2}), we only keep the most relevant operators
around the FM fixed point, described by ${\mathcal L}_0$, in the
sense of RG, and $\bm{j}_h(\bm{j}_b )$ and $\rho_h(\rho_b)$ are the
current and density operators for the $\tilde{h}$ ($\tilde{b}$)
fields, respectively.

Since ${\mathcal L}_{int}$ is irrelevant around the fixed point
described by ${\mathcal L}_0$, our low energy effective theory
shows clearly that the FM phase of the QDE model exhibits the
phenomenon of spin-charge separation. That is, the low energy
excitations are holons described by the $\tilde{h}$ field, which
carry charge $e$ and spin zero, and the FM spin waves (magnons)
described by the $\tilde{b}$ field, which are charge neutral and
carry spin-$1$. This is very different from the usual itinerant
ferromagnets, where the elementary excitations are the dressed
electrons, which carry charge $-e$ and spin-$1/2$, and the
magnons, which are collective excitations in the particle-hole
channel.

\section{Conclusion and discussions}

To summarize, we have studied the low-energy physics of the QDE
model based on the slave fermion formulation. The most important
feature of this approach is that the effects of large values of
$U$ and $J_H$ are taken into account right from the beginning.
(For the one-orbital model, the large $U,J_H$ limit is implemented
by the NDO condition.) This results in a non-Fermi-liquid ground
state, in contrast to most of the previous studies on the DE
model. A direct consequence following it is that the electron
spectral function exhibits a broad asymmetric peak away from the
Fermi surface, in addition to the quasiparticle peak. Both our
results about the low energy magnetic properties and
low-temperature thermodynamics are also consistent with
experimental data in the FM metallic phase. On the other hand, our
prediction for the optical conductivity at low energy is only
qualitatively consistent with experiments. Quantitatively, the
Drude weight we obtained is larger than that actually observed.
This indicates that to have a good quantitative description for
the low-frequency optical spectra of manganties, some extra
ingredients beyond the one-orbital QDE model must be taken into
account. However, we should emphasize that, for such a scenario to
be valid, the inclusion of these new ingredients should not affect
the low-temperature thermodynamics and low-energy magnetic
properties predicted by the one-orbital QDE model in any drastic
way because the one-orbital model already gives rise to reasonable
results on them.

We have also studied the role of gauge fluctuations and derived a
low energy continuum effective theory. A previous study in terms
of the slave-fermion gauge theory claimed that the gapless {\it
longitudinal} gauge fluctuations play a dominant role on the
electronic spectral properties.\cite{Hu} However, the longitudinal
component of gauge fields is gauge dependent and physics should be
independent of the gauge choice. Moreover, both the time and
longitudinal components of gauge fields are screened in the
metallic phase, which cannot affect low energy physics
dramatically. What we do in Sec. \ref{gauge} is to explicitly show
that the gauge fluctuations in the FM phase, both the transverse
and longitudinal components, are completely screened due to the
Anderson-Higgs mechanism. The resulting electron spectral function
in the FM metallic phase consists of two parts: one the sharp
quasiparticle peak at the Fermi energy and the other an asymmetric
broad peak away from the Fermi energy, instead of a broad
quasiparticle peak at the Fermi energy. These predictions can be
observed from the ARPES or STS measurements.

From the theoretical point of view, the slave fermion formulation
also gives a clear picture of the apparently observed Fermi liquid
behavior and its possible deviations. Within this framework, the
Fermi-liquid-like behaviors, such as a small value of the Drude
weight in the optical spectra and a small coherent quasi-particle
peak in the electron spectral function, are consequences of the
Bose condensation of the spinon field. Therefore, the spectral
weight is directly related to the magnitude of the condensate.
Based on the similar reasoning, we expect that in low dimensional
or layered materials where the quantum fluctuations tend to kill
the Bose condensate, the coherent quasi-particle peak will either
disappear completely or be greatly reduced in comparison with the
three-dimensional results we obtained here.\cite{Kaplan2,
Dagotto2} This is actually observed in the ARPES measurements for
layered compounds.\cite{Dessau,Chuang} Another minimal extension
of the one-orbital QDE model, which has the potential to reduce
the quasi-particle peak or the Drude weight, is to include the
orbital degrees of freedom. In this case, we have two slave boson
fields: one describing the spin fluctuations and the other
describing the orbital fluctuations. If only the spin slave field
condenses at low temperature, the strong orbital fluctuations will
also tend to reduce the coherent parts in the optical conductivity
and the electron spectral function. Researches along these
directions are in progress.

\acknowledgments

We would like to thank C.-D. Hu for introducing Refs.
\onlinecite{Khaliullin2} and \onlinecite{Weisse} to us. We are also
grateful to the hospitality of National Center for Theoretical
Sciences (North), where this work was initiated. The work of Y.L.L
is supported by the National Science Council of Taiwan under Grant
No. NSC 95-2112-M-018-008. The work of Y.-W.L is supported by the
National Science Council of Taiwan under Grant No. NSC
95-2112-M-029-009.


\begin{thebibliography}{00}
\bibitem{foot1} For the one-orbital model, the large $J_H$ limit
amounts to the large $U$ limit, provided one can treat the NDO
condition exactly. An intuitive arguement goes as follows: The
large $J_H$ limit forces the spins of the conduction electrons to
be parallel to the core spins. In the FM phase, it follows that
there can be no doubly occupied sites due to the Pauli exclusion
principle. However, in our approach, we first impose the NDO
condition by taking the $U/t\rightarrow +\infty$ limit. This, in
our opinion, makes the introduction of Eq. (\ref{sb3}) clearer and
the derivation of our working Hamiltonian [Eq. (\ref{qdeh2})]
smoother. Furthermore, this approach can be easily generalized to
the slave-fermion formulation of the DE model with degenerate
orbitals.
\end{thebibliography}
\end{document}